%% file: photo-sys-main.tex
\documentclass[sigconf,table]{acmart}
\usepackage{xcolor}
\usepackage{booktabs} 

\setcopyright{rightsretained}

\acmConference[RecSys 2018]{ACM Conference on Recommender Systems}{October 2018}{Vancouver, BC Canada}
\acmYear{2018}
\copyrightyear{2018}

\makeatletter
\def\@copyrightspace{\relax}
\makeatother

\begin{document}
\title{Aesthetic Features for Personalized Photo Recommendation}

\author{Yu Qing Zhou}
\affiliation{%
  \institution{Shopify}
  }
\email{ivan.zhou@shopify.com}
\author{Ga Wu}
\affiliation{%
  \institution{University of Toronto}
 }
\email{wuga@mie.utoronto.ca}
\author{Scott Sanner}
\affiliation{%
  \institution{University of Toronto}
 }
\email{ssanner@mie.utoronto.ca}
\author{Putra Manggala}
\affiliation{%
  \institution{Shopify}
  }
\email{putra.manggala@shopify.com}

\begin{abstract}
Many photography websites such as Flickr, 500px, Unsplash, and Adobe Behance are used by amateur and professional photography enthusiasts. Unlike content-based image search, such users of photography websites are not just looking for photos with certain content, but more generally for photos with a certain photographic "aesthetic".  In this context, we explore personalized photo recommendation and propose two aesthetic feature extraction methods based on (i) color space and (ii) deep style transfer embeddings.  Using a dataset from 500px, we evaluate how these features can be best leveraged by collaborative filtering methods and show that (ii) provides a significant boost in photo recommendation performance.
\end{abstract}


\maketitle

\input{photo-sys-body}
\bibliographystyle{ACM-Reference-Format}
\bibliography{photo-sys}

\end{document}

%% file: photo-sys-body.tex
\section{Introduction}


Personalized aesthetic-based photo recommendations can help
photography websites better serve the needs of amateur and professional photographers.
Content-based image search does not fully satisfy the needs of such users since they are usually not interested in content alone.  Instead, they are often looking for photos with certain photographic aesthetics, which may include monochromaticity, light contrast, and style.  While there has been research on automatic assessment of image aesthetics (e.g., \cite{Datta2006StudyingApproach, Deng2017ImageSurvey}) and clothing recommendation leveraging aesthetic features (e.g., most recently 
\cite{Yu2018Aesthetic-basedRecommendation}),
this work is not appropriate for personalized photo recommendation for photography enthusiasts as we detail in Section~2.

In this paper, we conduct experiments on a dataset from 500px.  We propose two feature extraction approaches to obtain aesthetic features from photos without manual annotation. The first approach extracts features from the Hue-Saturation-Value (HSV) channels of a photo since these align with human perception of brightness and whiteness~\citep{Cheng2001ColorProspects}. The second approach extracts deep style embeddings from photos~\cite{Gatys2016ImageNetworks}. Empirically, we show that the use of deep style embeddings as side information outperforms a variety of baselines. 

\section{Related Work}	

Many existing works in the literature on image aesthetic assessment are based on Photo.Net~\citep{Datta2006StudyingApproach} or datasets from the DPChallenge, like AVA ~\citep{Deng2017ImageSurvey}.
These datasets were annotated with semantic and aesthetic labels and rated by users unidentifiable to researchers.
For the purposes of this paper, it is not clear that these annotators align with the photography enthusiast community.  Further, these works explored (non-personalized) image classification tasks.

Some previous work has leveraged stylistic and aesthetic image features for personalized recommendation 
of fashion products. For example, \citep{McAuley2015Image-BasedSubstitutes} used style-based features derived from deep embeddings for personalized clothing recommendation.
Alternatively, \citep{Yu2018Aesthetic-basedRecommendation} incorporated 
simple binary aesthetic features (e.g., \textit{if the images are aesthetically pleasing to the public or not}).
In contrast, we focus on extracting rich aesthetic photo features such as color composition and texture-based deep style embeddings of~\citep{Gatys2016ImageNetworks} that we conjecture may better relate to photography enthusiast preferences.



\section{Approach}
\label{sec:approach}

Photo recommendation can be formalized in the usual matrix view: given a set of $m$ users, a set of $n$ photos, and the observed interaction of users with photos $R\in \{0, 1\}$ (1 indicates a positive interaction, and 0 indicates no observed interaction) with shape of $m \times n$, we want to rank the top-$k$ images that a user may positively interact with in the future.

One variant\footnote{In another variant of I-NN, the predicted interaction in~\eqref{eq:main_pred} is normalized by $\sum_{k \in \{ 1 \ldots n, k \neq j \}} \mathit{Sim}(\phi(j),\phi(k))$.  We observe better ranking performance without this.} of  item-item Nearest Neighbor Collaborative Filtering (I-NN)~\citep{Linden2003Amazon.comFiltering} predicts user $i$'s interaction with photo $j$ 
as 
\begin{equation} 
\label{eq:main_pred}
\hat{r}_{i,j} = \sum_{k \in \{ 1 \ldots n, k \neq j \}} \mathit{Sim}(\phi(j),\phi(k)) \cdot r_{i,k} ,
\end{equation}
where $\phi(j)$ is a vector of information for item $j$, and $\mathit{Sim}(\phi(j),\phi(k))$ defines similarity between photos $j$ and $k$; multiple similarity functions $\mathit{Sim}$ are available such as Cosine, Pearson, and Euclidean. If there is side information $\mathbf{p}_j$ available, in addition to the rating column $\mathbf{r}_{:,j}$ in matrix $R$, it is  expressed as
\begin{equation}\label{eq:itemsim}
Sim(\phi(j), \phi(k))= \theta \cdot Sim(\mathbf{p}_j, \mathbf{p}_k) + (1-\theta)\cdot Sim(\mathbf{r}_{:,j}, \mathbf{r}_{:,k}),
\end{equation}
where $\theta$ is a relative weighting hyperparameter that can be tuned through cross-validation. 

Next we describe two aesthetic feature extraction methods.



\subsubsection*{HSV Color-Embedding}
\begin{figure}
  \centering
\includegraphics[width=2.3cm, height=1.7cm]{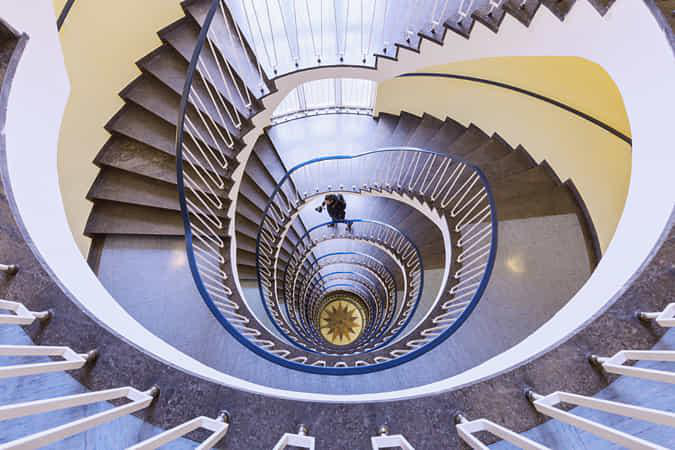}
\includegraphics[width=6.1cm, height=2.4cm]{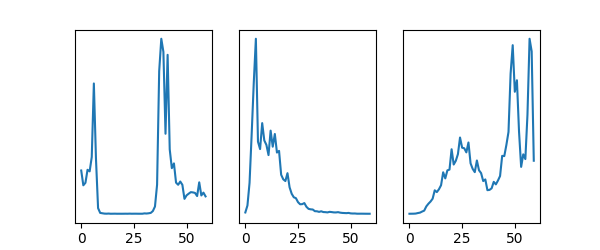}
\caption[]{(left) An example photo. (right) Density plot of Hue, Saturation, and Value channels for the photo.}
\label{fig:hsv_example}
\end{figure}

The HSV color space was designed to capture the human perception of color~\citep{Cheng2001ColorProspects}. It has been used in (non-personalized) photo aesthetic assessment \citep{Datta2006StudyingApproach}. In contrast with the traditional RGB color space, HSV separates out luminance from color information. 
We represent the HSV (or RGB) color-embedding $\mathbf{p}_j$ of photo $j$ as a concatenated histogram vector of the three HSV (or RGB) channels illustrated in Figure~\ref{fig:hsv_example}.


\subsubsection*{Style Embedding}
Histograms extracted from HSV or RGB channels are intuitive, but fail to capture more complex photo properties such as style (e.g., textures).
Fortunately, recent research in neural style transfer~\citep{Gatys2016ImageNetworks} shows that 
stylistic information of an image can be extracted 
with a deep convolutional neural network.
Specifically, the style embedding $\mathbf{p}_j$ is the vectorized form of the Gram matrix of a specific layer (determined through experimentation) from an inference performed on the photo $j$ using a VGG-19 network \cite{Simonyan2014VeryRecognitionb}.


\section{Results} \label{results}
We conduct our experiments on a dataset from 500px, an online photography website. The dataset contains 225,922 users and 300,000 photos. 
We prepared five temporally-split triples of train, validation, and test sets from this dataset. The validation set is used to tune hyperparameters, which are $\theta$ in~\eqref{eq:itemsim} and the similarity measure (either Cosine, Euclidean, or Pearson).  We evaluate three different ranking metrics:
\textit{Precision@10}, \textit{R-Precision}, and \textit{Average Precision}.

The baseline models used in the experiment are the following:
\begin{enumerate}
\item \textit{Random}: photos are ranked in a random order. 
\item \textit{Popular}: photos are ranked by popularity (most likes).
\item \textit{I-NN}: item-based nearest neighbour as covered in Section~\ref{sec:approach}.
\item \textit{I-NN-Meta}: \textit{I-NN} with photo metadata ($\mathbf{p}_j$), which includes categories, keywords, and \textit{"Editor's Choice"} labels. 
\end{enumerate}
The novel photo recommendation models (including typical hyperparameter settings from validation-based tuning) that we compare against the baselines are \textit{I-NN-HSV} ($\theta=0.01$, Cosine), 
\textit{I-NN-RGB} ($\theta=0.04$, Cosine), and \textit{I-NN-Style} ($\theta=0.20$ weights the sum of Euclidean on style vectors and Cosine on interaction vectors), which use aesthetic features from the HSV and RGB color channels, and deep style embeddings~\citep{Gatys2016ImageNetworks}, respectively, as defined in Section~\ref{sec:approach}.

Next, we experimentally answer a few research questions.

\vspace{1mm}
\noindent \textbf{Is HSV better than RGB?}  In Table~\ref{tab:visualresults}, we note that \textit{I-NN-HSV} outperforms \textit{I-NN-RGB} indicating that HSV color features may provide better similarity information for photo recommendation than RGB as we conjectured.  However, color information may not help at all since neither outperformed the best baseline \textit{I-NN}.



\begin{figure}[t!]
  \centering
\includegraphics[width=3.8cm, height=2.7cm]{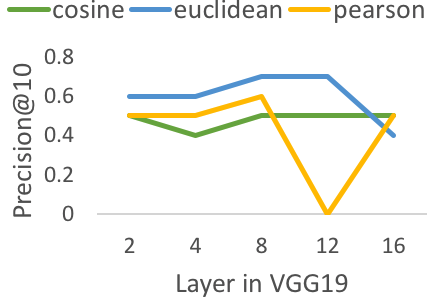}
\includegraphics[width=3.8cm, height=2.7cm]{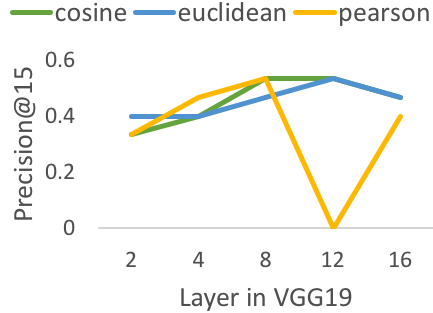}
\vspace{-1mm}
\caption[Optimal Configuration in Style Extraction]{Evaluation of VGG-19 layer and  metric for style extraction that maximizes recommendation performance.}
\label{fig:vgg_experiment}
\end{figure}

\vspace{1mm}
\noindent \textbf{What is the best layer of VGG and best distance metric for determining style-based similarity? } We set up an experiment with multiple sets of photos that are aesthetically similar to each other based on human annotation. We extract the style vectors from all photos with one specific layer of VGG-19 and then apply one of the similarity measures (Cosine, Euclidean, and Pearson) to compute the similarity between style vectors. We select the top $k$ pairs of photos based on similarity scores and determine how many of these pairs were in the same aesthetic set (denoted as the \textit{Precision@k} for $k=10$ and $k=15$). The results are provided in Figure~\ref{fig:vgg_experiment}. The Euclidean similarity metric applied to style embeddings extracted from layer 8 of VGG-19 gives us the highest \textit{Precision@k} for both  $k=10$ and $k=15$.  Thus, we use this configuration for \textit{I-NN-Style}.

\vspace{1mm}
\noindent \textbf{Which photo recommender  performs best overall?}
Table~\ref{tab:visualresults} provides comparative results.  As noted earlier, \textit{I-NN} performs best among baselines and outperforms three methods that use photo-related information: \textit{I-NN-Meta}, \textit{I-NN-HSV}, and \textit{I-NN-RGB}.  However, the best performer overall by a substantial margin is \textit{I-NN-Style} that uses photo aesthetic information derived from style embeddings.

\section{Conclusion}
Our results show that color and explicit metadata side information for photos do not help photo recommendation performance.  However, style embeddings derived from layer 8 of VGG-19 provide a significant boost, demonstrating the importance of aesthetic style features in recommendation for photography enthusiasts.


\begin{table}[t!]
\centering
\caption[Performance - Style Feature-Embedded Model]{Recommender performance comparison ($\pm$95\% CIs).}
\label{tab:visualresults}
\resizebox{\linewidth}{!}{%
\rowcolors{2}{gray!10}{white}
\begin{tabular}{@{}lllll@{}}
\toprule
Model  & Precision@10 & R-Precision & Avg Precision\\
\midrule
Random & 0.006 $\pm$ 0.002 & 0.006 $\pm$ 0.002 & 0.006 $\pm$ 0.002\\
Popular & 0.021 $\pm$ 0.004 & 0.018 $\pm$ 0.004 & 0.021 $\pm$ 0.005\\
I-NN & 0.048 $\pm$ 0.009 & 0.041 $\pm$ 0.008 & 0.047 $\pm$ 0.010\\
I-NN-Meta & 0.044 $\pm$ 0.005 & 0.037 $\pm$ 0.004 & 0.042 $\pm$ 0.006\\
\midrule
I-NN-HSV  & 0.041 $\pm$ 0.006  & 0.033 $\pm$ 0.006 & 0.040 $\pm$ 0.008\\
I-NN-RGB & 0.038 $\pm$ 0.003  & 0.030 $\pm$ 0.004 & 0.037 $\pm$ 0.006\\
I-NN-Style & \textbf{0.059 $\pm$ 0.012} & \textbf{0.050 $\pm$ 0.011} & \textbf{0.057 $\pm$ 0.014}\\
\bottomrule
\end{tabular}
}
\end{table}

%


%% file: photo-sys-main.bbl

\begin{thebibliography}{8}


\ifx \showCODEN    \undefined \def \showCODEN     #1{\unskip}     \fi
\ifx \showDOI      \undefined \def \showDOI       #1{#1}\fi
\ifx \showISBNx    \undefined \def \showISBNx     #1{\unskip}     \fi
\ifx \showISBNxiii \undefined \def \showISBNxiii  #1{\unskip}     \fi
\ifx \showISSN     \undefined \def \showISSN      #1{\unskip}     \fi
\ifx \showLCCN     \undefined \def \showLCCN      #1{\unskip}     \fi
\ifx \shownote     \undefined \def \shownote      #1{#1}          \fi
\ifx \showarticletitle \undefined \def \showarticletitle #1{#1}   \fi
\ifx \showURL      \undefined \def \showURL       {\relax}        \fi
\providecommand\bibfield[2]{#2}
\providecommand\bibinfo[2]{#2}
\providecommand\natexlab[1]{#1}
\providecommand\showeprint[2][]{arXiv:#2}

\bibitem[\protect\citeauthoryear{Cheng, Jiang, Sun, and Wang}{Cheng
  et~al\mbox{.}}{2001}]%
        {Cheng2001ColorProspects}
\bibfield{author}{\bibinfo{person}{H.D. Cheng}, \bibinfo{person}{X.H. Jiang},
  \bibinfo{person}{Y. Sun}, {and} \bibinfo{person}{Jingli Wang}.}
  \bibinfo{year}{2001}\natexlab{}.
\newblock \showarticletitle{{Color image segmentation: advances and
  prospects}}.
\newblock \bibinfo{journal}{\emph{Pattern Recognition}} \bibinfo{volume}{34},
  \bibinfo{number}{12} (\bibinfo{date}{12} \bibinfo{year}{2001}),
  \bibinfo{pages}{2259--2281}.
\newblock


\bibitem[\protect\citeauthoryear{Datta, Joshi, Li, and Wang}{Datta
  et~al\mbox{.}}{2006}]%
        {Datta2006StudyingApproach}
\bibfield{author}{\bibinfo{person}{Ritendra Datta}, \bibinfo{person}{Dhiraj
  Joshi}, \bibinfo{person}{Jia Li}, {and} \bibinfo{person}{James~Z. Wang}.}
  \bibinfo{year}{2006}\natexlab{}.
\newblock \showarticletitle{{Studying Aesthetics in Photographic Images Using a
  Computational Approach}}.
\newblock \bibinfo{publisher}{Springer}, \bibinfo{pages}{288--301}.
\newblock


\bibitem[\protect\citeauthoryear{Deng, Loy, and Tang}{Deng
  et~al\mbox{.}}{2017}]%
        {Deng2017ImageSurvey}
\bibfield{author}{\bibinfo{person}{Yubin Deng}, \bibinfo{person}{Chen~Change
  Loy}, {and} \bibinfo{person}{Xiaoou Tang}.} \bibinfo{year}{2017}\natexlab{}.
\newblock \showarticletitle{{Image Aesthetic Assessment: An experimental
  survey}}.
\newblock \bibinfo{journal}{\emph{IEEE Signal Processing Magazine}}
  \bibinfo{volume}{34}, \bibinfo{number}{4} (\bibinfo{date}{7}
  \bibinfo{year}{2017}), \bibinfo{pages}{80--106}.
\newblock
\urldef\tempurl%
\url{http://ieeexplore.ieee.org/document/7974874/}
\showURL{%
\tempurl}


\bibitem[\protect\citeauthoryear{Gatys, Ecker, and Bethge}{Gatys
  et~al\mbox{.}}{2016}]%
        {Gatys2016ImageNetworks}
\bibfield{author}{\bibinfo{person}{Leon~A. Gatys},
  \bibinfo{person}{Alexander~S. Ecker}, {and} \bibinfo{person}{Matthias
  Bethge}.} \bibinfo{year}{2016}\natexlab{}.
\newblock \showarticletitle{{Image Style Transfer Using Convolutional Neural
  Networks}}. In \bibinfo{booktitle}{\emph{2016 CVPR}}.
  \bibinfo{publisher}{IEEE}, \bibinfo{pages}{2414--2423}.
\newblock


\bibitem[\protect\citeauthoryear{Linden, Smith, and York}{Linden
  et~al\mbox{.}}{2003}]%
        {Linden2003Amazon.comFiltering}
\bibfield{author}{\bibinfo{person}{G. Linden}, \bibinfo{person}{B. Smith},
  {and} \bibinfo{person}{J. York}.} \bibinfo{year}{2003}\natexlab{}.
\newblock \showarticletitle{{Amazon.com recommendations: item-to-item
  collaborative filtering}}.
\newblock \bibinfo{journal}{\emph{IEEE Internet Computing}}
  \bibinfo{volume}{7}, \bibinfo{number}{1} (\bibinfo{date}{1}
  \bibinfo{year}{2003}), \bibinfo{pages}{76--80}.
\newblock


\bibitem[\protect\citeauthoryear{McAuley, Targett, Shi, and van~den
  Hengel}{McAuley et~al\mbox{.}}{2015}]%
        {McAuley2015Image-BasedSubstitutes}
\bibfield{author}{\bibinfo{person}{Julian McAuley},
  \bibinfo{person}{Christopher Targett}, \bibinfo{person}{Qinfeng Shi}, {and}
  \bibinfo{person}{Anton van~den Hengel}.} \bibinfo{year}{2015}\natexlab{}.
\newblock \showarticletitle{{Image-Based Recommendations on Styles and
  Substitutes}}. In \bibinfo{booktitle}{\emph{Proceedings of SIGIR '15}}.
  \bibinfo{pages}{43--52}.
\newblock
\urldef\tempurl%
\url{http://dl.acm.org/citation.cfm?doid=2766462.2767755}
\showURL{%
\tempurl}


\bibitem[\protect\citeauthoryear{Simonyan and Zisserman}{Simonyan and
  Zisserman}{2014}]%
        {Simonyan2014VeryRecognitionb}
\bibfield{author}{\bibinfo{person}{Karen Simonyan} {and}
  \bibinfo{person}{Andrew Zisserman}.} \bibinfo{year}{2014}\natexlab{}.
\newblock \showarticletitle{{Very Deep Convolutional Networks for Large-Scale
  Image Recognition}}.
\newblock  (\bibinfo{date}{9} \bibinfo{year}{2014}).
\newblock
\urldef\tempurl%
\url{http://arxiv.org/abs/1409.1556}
\showURL{%
\tempurl}


\bibitem[\protect\citeauthoryear{Yu, Zhang, He, Chen, Xiong, and Qin}{Yu
  et~al\mbox{.}}{2018}]%
        {Yu2018Aesthetic-basedRecommendation}
\bibfield{author}{\bibinfo{person}{Wenhui Yu}, \bibinfo{person}{Huidi Zhang},
  \bibinfo{person}{Xiangnan He}, \bibinfo{person}{Xu Chen}, \bibinfo{person}{Li
  Xiong}, {and} \bibinfo{person}{Zheng Qin}.} \bibinfo{year}{2018}\natexlab{}.
\newblock \showarticletitle{{Aesthetic-based Clothing Recommendation}}. In
  \bibinfo{booktitle}{\emph{Proceedings of WWW '18}}.
  \bibinfo{pages}{649--658}.
\newblock


\end{thebibliography}
